\documentclass{procpavia}

\begin{document}
\pagestyle{prochead}

\topmargin -.4cm   


\title{Rescattering Contributions to Final State Interactions \\
          in $(e,e'p)$ Reactions at High $(p_m,E_m)$}
\author{C.~Barbieri}
  \email{barbieri@triumf.ca}
  \homepage{http://www.triumf.ca/people/barbieri}
  \affiliation
     {TRIUMF, 4004 Wesbrook Mall, Vancouver, 
          British Columbia, Canada V6T 2A3 \\  }

\author{L.~Lapik\'{a}s}
  \affiliation
     {NIKHEF, P.O. Box 41882, 1009 DB Amsterdam, The Netherlands \\  }

\begin{abstract}
The contribution of rescattering to  final state interactions
in the $(e,e'p)$ cross section is studied using a semiclassical model.
This approach considers a two-step process
with the propagation of an intermediate nucleon and uses Glauber theory
to account for the reduction of the experimental yield
due to N--N scattering.
 This calculation has relevance for the analysis of data
at high missing energies and in particular at the kinematics of
the E97-006 experiment done at JLab.
 It is found that rescattering is strongly reduced in parallel
kinematics and that the excitation of nucleon resonances
is likely to give important contributions to the final state interactions
in the correlated region.
 For heavy nuclei, further enhancement to the rescattering is expected to be
generated from the strength in the mean field region.
\end{abstract}

\noindent
\begin{flushright}
{TRIUMF preprint: TRI-PP-03-39}
\\
\end{flushright}

\maketitle
\setcounter{page}{1}
\preprint{TRI-PP-03-39}


\section{Introduction}

 Nuclear correlations strongly influence the dynamics
of nuclear systems. In particular, the repulsive core at small 
internucleon distances has the effect of removing the nucleons from
their shell model orbitals, producing pairs of nucleons with high
and opposite relative momenta. 
 This results in spreading out a sizable amount of spectral strength, 
about 10-15\%~\cite{bv91}, to very high missing energies and momenta and
in increasing the binding energy of the system~\cite{Wim03}. 
 This reduction appears to be fairly independent of the given subshell
and on the size of the nucleus, except for a slight increase with
the central density of the systems~\cite{WimPavia}.
 This is also in line with the existence of two nucleon substructures,
similar to the deuteron and independent on the atomic mass,
that have been discussed in Ref.~\cite{SchiavillaSR}
using the variational approach and have been found to be driven by
short-range and tensor correlations.
 Theoretical studies of the distribution of short-range correlated 
nucleons for finite nuclei have been carried out in Ref.~\cite{WHA}
and by Benhar et al.~\cite{Benhar}. These calculations suggest that most
of this strength is found along a ridge in the momentum-energy plane
($k$-$E$) which spans several hundreds of MeV/c (and MeV). The most probable
energy is the one of a free moving nucleon but shifted by a constant
term that represents the average two-nucleon separation energy for
that pair.
It is important to note that the depletion of single particle orbitals
observed near the Fermi energy is more substantial than
the 15\% reduction discussed above~\cite{Louk93}. This is mostly due
to long-range effects and cannot be understood in terms of short-range
correlations alone~\cite{WimPavia}.
 A proper description of the spectral function at low missing energies
usually requires the coupling of the single particle motion to low-energy
collective modes (see for example Ref.~\cite{LRC}).
However, a proper knowledge of short-range effects
remains of great importance for understanding of nuclear
systems. In particular the details of the strength distribution
at high missing energy strongly influence the binding energy
of finite nuclei and nuclear matter~\cite{Wim03}.
Hence, the importance determining this distribution experimentally.

\begin{figure}[t]
  \begin{center}
    \includegraphics[height=0.28\textheight]{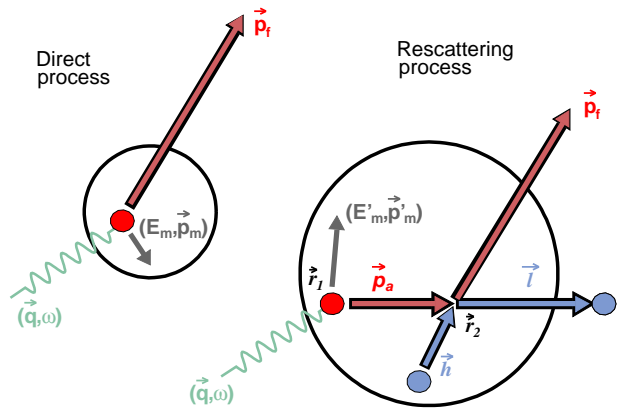}
    \caption{ \label{fig:TotRes}
       Schematic representation of the direct knockout of a proton,
      given by the PWIA (left) and the contribution from a two-step
      rescattering (right). In the latter a proton or neutron is emitted
      with momentum $\vec{p}_a$ and different missing energy and momentum
      $(E_m',\vec{p}_m')$. Due to a successive collision, a proton
      is finally detected with the same momentum $\vec{p}_f$ seen in the
      direct process.
  }
    \end{center}
\end{figure}

Signatures of correlations have been
found in the ${}^{16}{\rm O}(e,e'pp)$ reaction, where it has been seen 
that the transition to the ground state of ${}^{14}{\rm C}$
is dominated by short-range effects~\cite{O16pp}.
 Two-nucleon emission reactions have the capability of probing
the correlated pair directly, therefore obtaining information on
short-range correlation already at low missing energies~\cite{Boffi}.
 Moreover the effects of collective excitations and 
final state interactions (FSI) represent other important ingredients
to which the two-nucleon emission cross sections are
sensitive~\cite{GeurtsO16pp,HerbO16pn,schwamb}.
 Given the richness of information that can be accessed,
one can expect that further studies of two-nucleon emission
--both theoretically and experimentally-- are likely have
an important impact on the understanding of nuclear structure.

For a long time $(e,e'p)$ reactions have provided a useful tool
to determine the nuclear spectral functions at small missing energy. 
However, past measurements in the region of interest to short-range
correlations have been limited due to the enormous background
that is generated  by final state interactions, see for
example Ref.~\cite{MIT-BATES}.
The issue of how to minimize the FSI has been addressed
in Ref.~\cite{E97proposal}. There, it is shown that FSI
in exclusive $(e,e'p)$
cross sections are dominated by two-step processes like the one
depicted in Fig.~\ref{fig:TotRes}. This becomes particularly 
relevant when perpendicular kinematics are employed to probe
the regions of small spectral strength.
Other important contributions are expected to come from $(e,e'\Delta)$
reactions followed by the decay of the $\Delta$ resonance.
A study of the kinematic conditions shows that for  perpendicular
kinematics the rescattered nucleons move spectral strength in the
$k$-$E$ plane, from the top of the ridge toward regions where the
correlated strength is small, therefore submerging the direct signal
in a large  background noise.
 In Ref.~\cite{E97proposal}, it was suggested that the 
contribution of rescattering can be diminished in parallel kinematics.
 New data was subsequently taken
in these conditions at Jefferson Lab~\cite{danielaElba,danielaPavia}.
We also note that a similar trend of the FSI with respect to the kinematics
is also in line with the prediction of Ref.~\cite{MuetherFSI}
for $(e,e'NN)$ reaction in superparallel kinematics.

The theoretical calculation of the rescattering yield has been
addressed in Ref.~\cite{Marcel}. This calculation was based on the
semiclassical model of Ref.~\cite{VjPi} and intended to be 
applied at lower missing energies. This contribution reports
about an ongoing work aimed to extend these calculations
to the kinematics of interest for the study of short-range
correlations. The model of Ref.~\cite{Marcel} is discussed
and extended to a relativistic notation in Sec.~\ref{sec2}.
First results for the kinematics of the
Jefferson Lab data are shown in Sec.~\ref{sec:results}.


\section{Model}
\label{sec2}

At large $E_m$ appreciable contributions to the experimental yield come
from two-step
mechanisms, in which a reaction $(e,e'a)$ is followed by a scattering process
from a nucleon in the nuclear medium, $N'(a,p)N''$, eventually leading
to the
emission of the detected proton. In general, $a$ may represent
a nucleon or another possible intermediate particle, as a $\Delta$
excitation. In the
following we will also use the letter $a$ to label the possible open channels.

Following the semiclassical approach proposed in Refs.~\cite{VjPi,Marcel},
the contribution to the cross section coming from rescattering
through the channel $a$ is written as
\begin{eqnarray}
    { d^6 ~ \sigma^{(a)}_{rescat}
     \over
     dE_0 \; d\Omega_{\hat{k}_o}  dE_f \; d \Omega_{\hat{p}_f} } &=&  
    \int_V d \vec{r}_1 \int_V d \vec{r}_2 \int_{0}^{\omega} d T_a
  \rho_N(\vec{r}_1) \; 
    { K \; S^h_N(p'_m,E'_m) \; \sigma^{cc1}_{eN}
    \over
    A \; ( \vec{r}_1 - \vec{r}_2 )^2 } 
    g_{aN'}(|\vec{r}_1 - \vec{r}_2|) \; \; \;
\nonumber  \\
 &\;& \times 
 P_T(p_a, \vec{r}_1 , \vec{r}_2 )
\rho_{N'}(\vec{r}_2) \; 
    { d^3 ~ \sigma_{a N'}
    \over
    dE_f \; d \Omega_{\hat{p}_f} } \;
P_T(p_f, \vec{r}_2 , \infty) \; ,
\label{eq:TotRes}
\end{eqnarray}
where $(E_o,\vec{k}_o)$ and  $(E_f,\vec{p}_f)$ represent the four-momenta
of the outgoing electron and proton,  respectively.
Eq.~(\ref{eq:TotRes}) assumes that the intermediate particle $a$ is
generated in PWIA by the electromagnetic current at a point $\vec{r}_1$
inside the nucleus. Here $K=|\vec{p}_a|E_a$ is a phase space factor,
$S^h_N(p'_m,E'_m)$ is
the spectral function of the hit particle $a$ and $\sigma^{cc1}_{eN}$
the off shell electron-nucleon cross section, for which
we have used the $cc1$ prescription of de~Forest~\cite{deForest}.
The pair distribution functions
$g_{aN'}(|\vec{r}_1 - \vec{r}_2|)$ that account for the joint probability of
finding a nucleon N' in $\vec{r}_2$ after the particle $a$ has been struck
at $\vec{r}_1$.
The kinetic energy $T_a$ of the intermediate particle $a$ is integrated
up to the energy $\omega$ transfered by the electron.
The transparency factor $P_T(p, \vec{r}_1 , \vec{r}_2)$ gives
the probability that the struck particle $a$ propagates to
a second point $\vec{r}_2$, where it scatters from the nucleon $N'$ with
cross section $d^3 ~ \sigma_{a N'}$. The whole process is depicted
in Fig.~\ref{fig:TotRes}.

In the calculation described in Sec.~\ref{sec:results} we will only
consider the channels in which $a$ is either a proton or a neutron.
It is clear that other channels are expected to be important.
In particular, the excitation of the $\Delta$ resonance is also seen 
to contribute from the preliminary data of the E97-006
experiment~\cite{danielaElba}.
  Moreover other channels involving the creation of pions are open at the 
kinematics considered here.

 Eq.~(\ref{eq:TotRes}) is a seven-fold integral that can be conveniently 
evaluated with Monte Carlo techniques, once the terms in the integrand
are known.
The following subsections describe  the calculation of the cross section
$d^3 ~ \sigma_{a N'}$ and of the transmission probability $P_T$.

\subsection{Evaluation of the in-medium nucleon-nucleon rate}

Eq.~(\ref{eq:TotRes}) requires the evaluation of the process
in which the particle $a$,
which could be either a proton or a neutron, hits against a bound
nucleon in its way out, eventually leading to the emission of the
detected proton.
For the present purposes the spectral distribution of the hit nucleon,
$N'$, can
be appropriately described by the free Fermi gas distribution.
The cross section is therefore computed for a nucleon $a$ travelling
in symmetric nuclear matter at a given density $\rho_{NM}$.
 The effects of the nuclear surface are eventually included
in Eq.~(\ref{eq:TotRes})
using the local density approximation, that is, by evaluating the
cross section for the density at the point $\vec{r}_2$.
Initially, the hit nucleon $N'$ is in the Fermi sea and therefore must have
a momentum $\vec{h}$ smaller than the Fermi momentum
$k_f = (3\pi^2\rho_{NM}/2)^{1/3}$.
 At the same time the Pauli principle requires that the
particles in the final state will have momenta $\vec{p}_f$ and $\vec{l}$,
both larger than $k_f$.
 Among all the nucleons involved in the process,
$\vec{p}_f$ will refer to the detected proton while the others can be
either neutrons or protons depending on the channel $a$.

The probability per unit time of an event leading to the emission
of a proton with momentum $\vec{p}_f$ is obtained by imposing the
Pauli constraints and integrating over the unobserved
momenta $\vec{h}$ and $\vec{l}$. Employing a relativistic
notation,
\begin{eqnarray}
d^3 P 
 \over dp_f d\Omega_{\hat{p}_f} &=&
  2 \; \theta(p_f - k_f)  \; L^3
      \int_{L^3} \int_{L^3} {d\vec{h} \;  d\vec{l} \over (2\pi)^6} 
      \; \theta(k_f - h) \theta(l - k_f) \;
         W_I
\nonumber \\
 &=& 2 \; p_f^2 \; \theta(p_f - k_f) \int_{L^3} 
                  {d\vec{h} \over (2\pi)^3} 
    \theta(k_f - h) \theta(l - k_f)
            { m_a \; m_h \over E_a(p_a) E_h(h)} 
\label{eq:P_dpf}  \\
  & & ~ ~ ~\left. 
   \times { \left| {\cal M}(s,t,u) \right|^2 \over  4 \pi^2}
    { m_p \; m_l \over E_p(p_f) E_l(l)} \delta(E_a + E_h - E_f - E_l)
 \right|_{\vec{l} = \vec{p}_a + \vec{h}_f - \vec{p}_f} \; ,
\nonumber 
\end{eqnarray}
where $L^3$ is the volume of a normalization box
and $E_N(p) = (p^2 + m_N^2)^{1/2}$.
%
%
In Eq.~(\ref{eq:P_dpf}), $W_I$ is the probability per unit time for
the event $p_a^\mu + h^\mu \rightarrow p_f^\mu + l^\mu$  which is
expressed in terms of the Lorentz 
invariant amplitude ${\cal M}(s,t,u)$~\cite{PeSch}.
 In the present work, it is assumed that this is not appreciably
modified by the in-medium effects and therefore the on-shell
values for $\left|{\cal M}(s,t,u)\right|^2$ were used~\cite{VjPi}.
 These were extracted from the
free scattering cross section generated by the SAID
phase shift data analysis~\cite{SAID}.

It should also be noted that Eq.~(\ref{eq:P_dpf}) is related to the inverse
life time of the nucleon $a$ by
\begin{equation}
 \frac{1}{\tau_a} = \int d\Omega_{\hat{p}_f} \int d p_f 
{d^3P \over dp_f d\Omega_{\hat{p}_f} }  \; ,
\end{equation}
from which the nuclear transparency can be evaluated~\cite{VjPi}.
The in medium scattering rate is finally related to Eq.~(\ref{eq:P_dpf}) by
\begin{equation}
  { d^3 ~ \sigma_{a N'}
    \over
    dE_f \; d \Omega_{\hat{p}_f} }
    = {E_a \over \rho_{N'} p_a }{ E_f \over p_f}
  { d^3 P   \over dp_f d\Omega_{\hat{p}_f} }
    \; .
 \label{eq:SaN_dpf}
\end{equation}

\subsection{Transparency factor}

 According to Glauber theory,  the
probability $P_T$ that a proton struck at $\vec{r}_1$ will travel
with momentum $\vec{p}$ to the point $\vec{r}_2$ without being
rescattered is given by
\begin{eqnarray}
P_T(p,\vec{r}_1,\vec{r}_2) = & 
exp \left\{ - \int_{z_1}^{z_2}  \right.dz & 
    \left[ g_{pp}(|\vec{r}_1 - \vec{r}|) \; \tilde{\sigma}_{pp}(p,\rho(\vec{r}))
           \; \rho_p(\vec{r})   \right.
\label{eq:PT} \\
   & &  \left. \left. ~+~  g_{pn}(|\vec{r}_1 - \vec{r}|) \; \tilde{\sigma}_{pn}(p,\rho(\vec{r}))
           \; \rho_n(\vec{r}) \right]  \right\} \; ,
\nonumber
\end{eqnarray}
where the z axis is chosen along the direction of propagation
$\vec{p}$, an impact
parameter $\vec{b}$ is defined so that $\vec{r} = \vec{b} + z \hat{p}$,
and $z_1$ ($z_2$) refer to the initial (final) position.
The in medium total cross sections $\tilde{\sigma}_{pp}(p,\rho)$ and
$\tilde{\sigma}_{pn}(p,\rho)$ have been computed in Ref.~\cite{VjPi}
up to energies of 300~MeV and account for the effects of Pauli blocking,
Fermi spreading and the velocity dependence of the nuclear mean field.
 For energies above 300~MeV these
have been extended to incorporate effects of pion emission~\cite{Pieperpriv}.
We note that Eq.~(\ref{eq:PT}) differs from the standard Glauber theory
by the inclusion of the pair distribution functions
$g_{pN}(|\vec{r}_1 - \vec{r}|)$.
 In principle, the $g_{pN}$ functions should depend on the
density and on the direction of the inter-particle distance.
 However, these
effects has been shown to be negligible in Ref.~\cite{VjPi}.
In the present application we find that a simple two-gaussian
parametrization of the $g_{pN}$ can adequately fit the curves
reported in Ref.~\cite{VjPi} for nuclear matter at saturation density.

The nuclear transparency is defined, in Glauber theory,
as the average over the nucleus of the probability that the struck proton
emerges from the nucleus without any collision. This is related
to $P_T$ by
\begin{equation}
 T = {1 \over Z} \int d\vec{r} \rho_p(\vec{r}) P_T(p,\vec{r},\infty) \; .
 \label{eq:T}
\end{equation}
in the case of an outgoing proton with
energy $E_f \sim$~1.8~GeV, which is of interest for the present
application, we find that $T =$~0.63 for ${}^{12}{\rm C}$ and $T =$~0.29
for ${}^{197}{\rm Au}$.


\begin{figure}[!t]
\vspace{.2in}
  \begin{center}
    \includegraphics[width=0.70\linewidth]{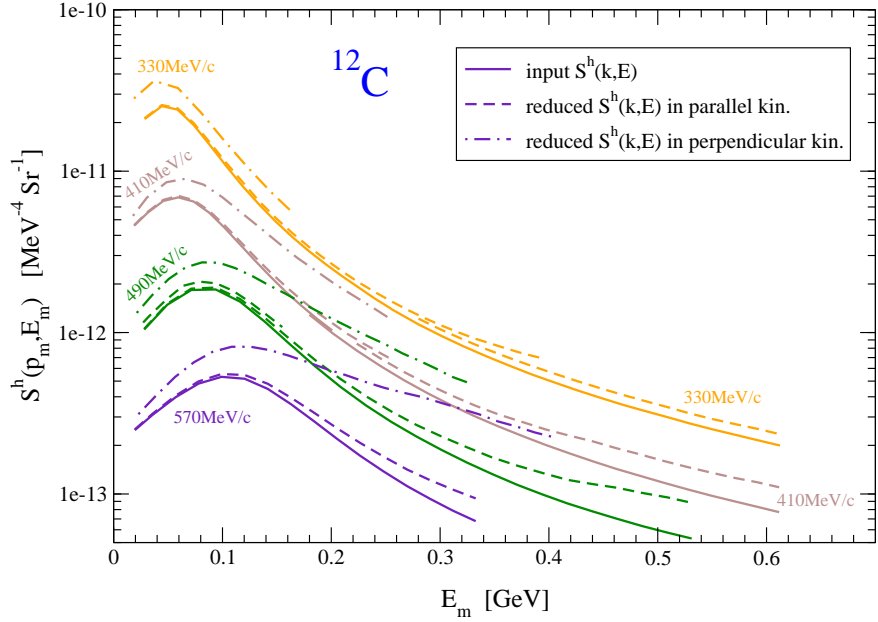}
    \caption{ \label{fig:res1}
      Theoretical results for the reduced spectral strength in the
     correlated region obtained in parallel (dashed line) and
     perpendicular (dot-dashed line) kinematics.  The full line shows
     the model spectral function, Eq.~(\ref{eq:Shtotal}), employed
     in the calculations. All lines refer to a ${}^{12}{\rm C}$ target.
     }
  \end{center}
\end{figure}

\section{Results}
\label{sec:results}

At energies close to the Fermi level the hole spectral function 
is dominated by its mean field component.
For ${}^{12}{\rm C}(e,e'p)$ these are orbitals
in the $s$ and $p$ shells, are know
experimentally and represent about 60~\% of the total
distribution~\cite{LoukC12}.
 It is therefore convenient to split the spectral function in a mean field
and a correlated part, 
\begin{equation}
 S^h_p(p_m,E_m) = S^h_{MF}(p_m,E_m) + S^h_{corr}(p_m,E_m) \; ,
 \label{eq:Shtotal}
\end{equation}
in which $S^h_{corr}(p_m,E_m)$ also contains the short-range correlated
tail at very high missing energies and momenta~\cite{WHA,Benhar}. 
In the following calculations we parametrize it as
\begin{equation}
 S^h_{corr}(p_m,E_m) = 
   { C \;  e^{- \alpha \, p_m} \over [E_m - e(p_m)]^2 + [\Gamma(p_m)/2]}
 \label{eq:Shcorr}
\end{equation}
where $e(p_m)$ and $\Gamma(p_m)$ are smooth functions of the missing 
momentum and the
parameters were chosen to give an appropriate fit to the available
${}^{12}{\rm C}(e,e'p)$ data in parallel kinematics~\cite{danielaPavia}.
The solid line in Fig.~\ref{fig:res1} shows the model spectral function,
Eq.~(\ref{eq:Shtotal}), employed in the present calculations
for that part of the  $k$-$E$ plane where $S^h_{corr}$ dominates.
The calculation with a gold target employed the same $S^h_{corr}$ of
Eq.~(\ref{eq:Shcorr}) multiplied by~$79/6$ to account for the
right number of protons. Since no direct data is available for the 
mean field spectral function of ${}^{197}{\rm Au}$ we choose to employ
the spectral function recently measured at NIKHEF for the neighbour
nucleus ${}^{208}{\rm Pb}$~\cite{Marcel},
also rescaled to the number of protons in gold.

We have performed calculations of the rescattering contribution by
employing both parallel and perpendicular kinematics.
 In the first case,
the angle  between the momentum transfered by the 
electron and the momentum of the final proton was chosen to be
$\vartheta_{qf} \sim$~5~deg
and the energy of the final proton was $E_f\sim$~1.6~GeV.
 For the perpendicular kinematics, $\vartheta_{qf}\sim$~30~deg
and $E_f\sim$~1~GeV. In both cases the four momentum transfered by
the electron was $Q^2 \sim$~0.40~GeV$^2$.

The results for the total cross section 
($\sigma_{PWIA}+\sigma_{rescat}$) have been converted to a reduced spectral 
function representation by dividing
them by $|p_f E_f|T\sigma_{eN}^{cc1}$,
evaluated for the kinematics of the direct process.
This is coherent with the analysis carried out for the experimental
data discussed below~\cite{danielaPavia}.
Fig.~\ref{fig:res1} shows the results from Eq.~(\ref{eq:TotRes})
for ${}^{12}{\rm C}$ in both kinematics.
 As can be seen, FSI from nucleon-nucleon rescattering give little
contribution to the total cross section in parallel kinematics, and
the resulting reduced spectral function is close to the true one.
For perpendicular kinematics, more sizable contributions are found
and they tend to fill the region at higher missing energies,
where the spectral function is small.
 This confirms the trend of FSI expected for parallel kinematics
that strength is primarly moved from places where  $S^h(p_m,E_m)$
is small to places where it is large, thus giving a small relative
effect~\cite{E97proposal}.
%

\begin{figure}[!t]
\vspace{.15in}
  \begin{center}
    \includegraphics[width=0.70\linewidth]{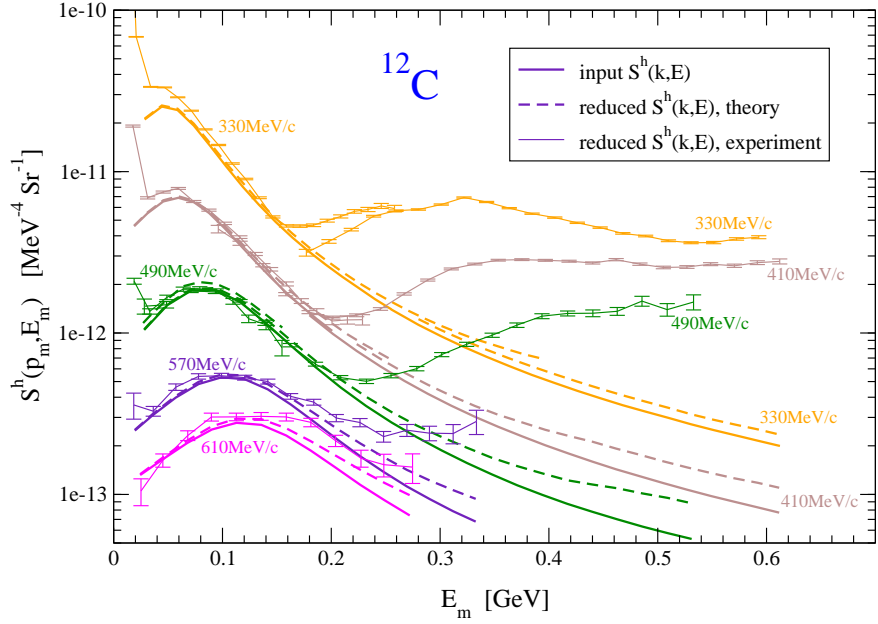}
    \caption{ \label{fig:res2}
       Theoretical results for the reduced spectral strength of
       ${}^{12}{\rm C}$  obtained in
      parallel kinematics (dashed line) compared to the experimental
      results of Ref.~\cite{danielaPavia}.
       The full line shows the model spectral function of
      Eq.~(\ref{eq:Shtotal}) employed in the calculations.
}
\end{center}
\end{figure}
\begin{figure}[!tb]
\vspace{.15in}
  \begin{center}
    \includegraphics[width=0.70\linewidth]{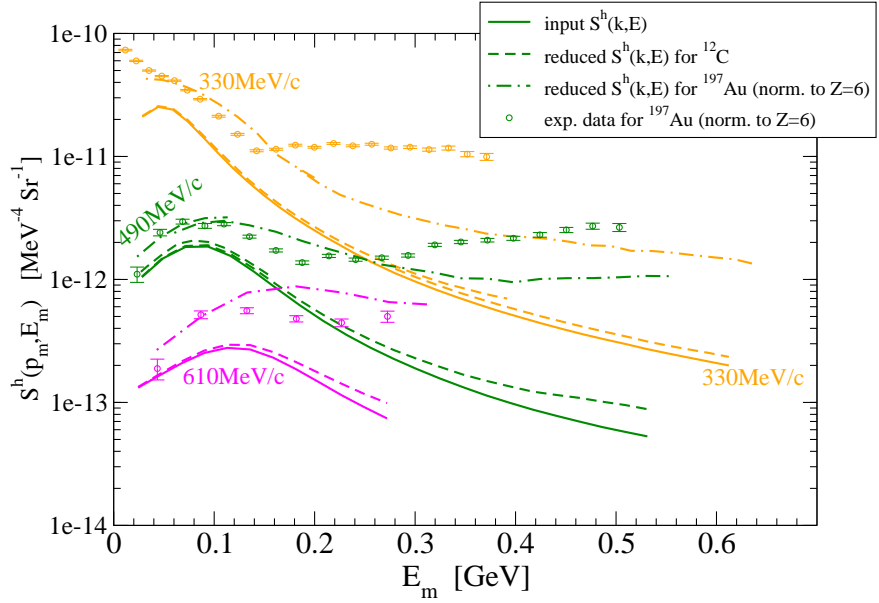}
    \caption{ \label{fig:res3}
       Theoretical results for the reduced spectral strength obtained in
      parallel kinematics for ${}^{197}{\rm Au}$ (dot-dashed line)
      compared to
      the relative experimental results of Ref.~\cite{danielaPavia}.
       The curves are  normalized to the same
      number of protons in ${}^{12}{\rm C}$ and compared with the
      theoretical results in parallel
      kinematics for this nucleus (dashed line).
       The full line shows the model spectral function of
      Eq.~(\ref{eq:Shtotal}) employed in the calculations.
}
\end{center}
\end{figure}

Figures~\ref{fig:res2} and~\ref{fig:res3} compare the model spectral
function~(\ref{eq:Shtotal}) and the theoretical reduced one,
with preliminary results from the E97-006 collaboration
in parallel kinematics~\cite{danielaPavia}.
 In both cases, the present model appears to be able to properly describe
the contribution to FSI in the correlated region.
Moreover Fig.~\ref{fig:res3} shows that the effects of rescattering are 
more relevant for heavier nuclei, if compared to the case of a small
nucleus like~${}^{12}{\rm C}$.
 An enhancement of the
cross section is found experimentally at very high missing energies and it
is presumably generated by the excitation of a $\Delta$ resonance.
 This effect
is not included in the present calculation yet. Contributions
from rescattering through this channel are expected to fill up the valley
between the correlated and $\Delta$ regions more substantially for heavier
nuclei.
 Therefore, these additional degrees of freedom need to be included
in the present model.

\begin{figure}[!t]
\vspace{.2in}
  \begin{center}
    \includegraphics[width=0.70\linewidth]{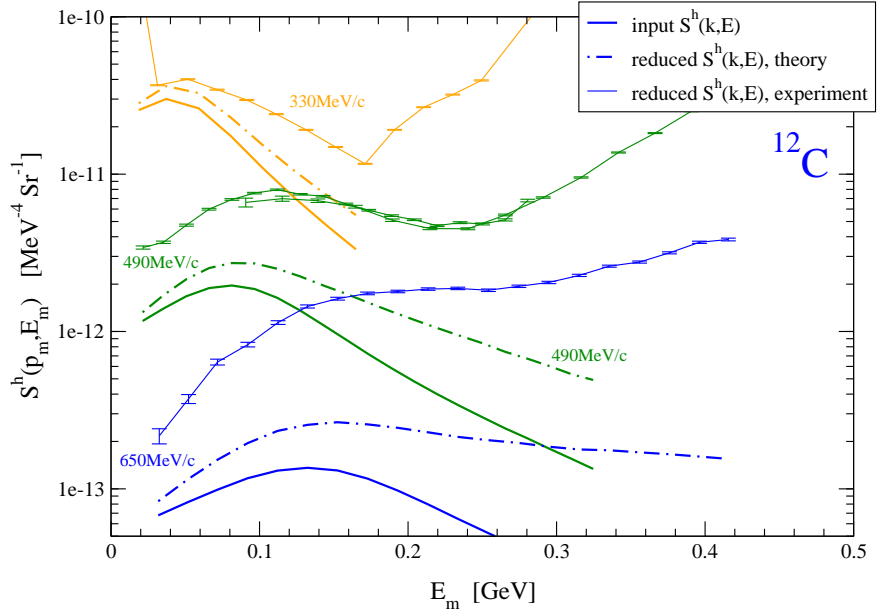}
    \caption{ \label{fig:res4}
       Theoretical results for the reduced spectral strength
       of ${}^{12}{\rm C}$ obtained in
      perpendicular kinematics (dot-dashed line) compared to the experimental
      results of Ref.~\cite{danielaPavia}.
       The full line shows the model spectral function of
      Eq.~(\ref{eq:Shtotal}) employed in the calculations.
}
\end{center}
\end{figure}

Figure~\ref{fig:res4} compares the present calculation for ${}^{12}{\rm C}$ 
to the experimental results in perpendicular kinematics.
 Although the present model explains the strong enhancement of rescattering 
with respect to the parallel case, see figure~\ref{fig:res1},
it significantly underestimates the experimental data.
 In this case the experimental data completely fill the valley between 
the correlated and $\Delta$ regions, suggesting that most of this
discrepancy is a consequence of neglecting this degree of freedom.
 Further contribution to the rescattering process can also come
form the inclusion of pion effects in the in-medium   cross
section, Eq.~(\ref{eq:SaN_dpf}). These effects may be added to 
Eq.~(\ref{eq:P_dpf}) by parametrrizing 
the invariant Lorentz amplitude ${\cal M}(s,t,u)$ in terms of the
total NN cross section~\cite{Ryckebusch03}.

\begin{figure}[t]
\vspace{.2in}
  \begin{center}
    \includegraphics[width=0.70\linewidth]{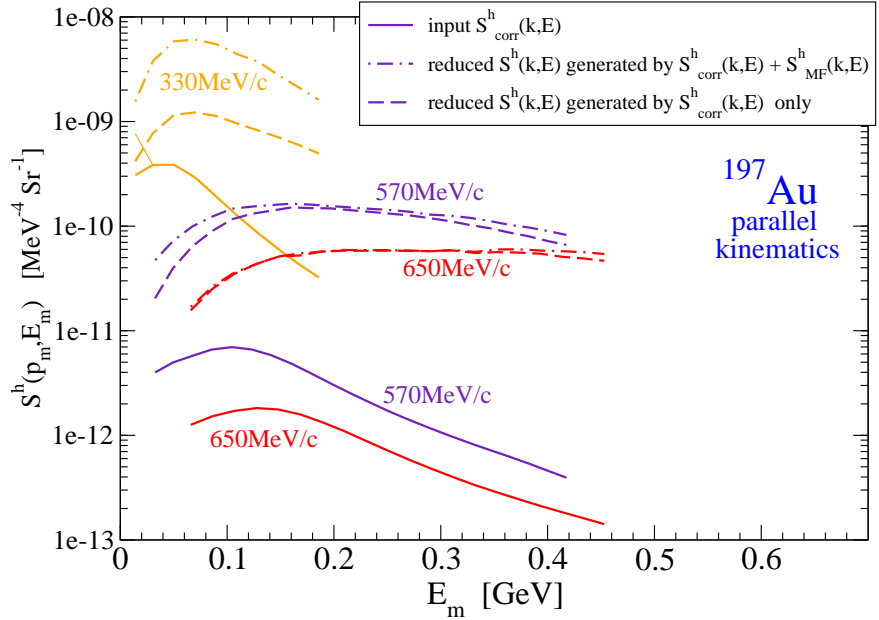}
    \caption{ \label{fig:res5}
       Reduced spectral strength of ${}^{197}{\rm Au}$ computed in
      parallel kinematics as generated by the full spectral function of
      Eq.~(\ref{eq:Shtotal}) (dot-dashed line) or by the sole correlated
      part $S^h_{corr}$ (dashed line).
       The full line shows the model spectral function $S^h_{corr}$
      employed in the calculations. The mean field part $S^h_{MF}$ is not
      visible this plot.
}
\end{center}
\end{figure}

In principle, one should expect that the  contribution
of rescattering increases as $A^{1/3}$ when moving to
heavier nuclei.
However, the experimental data suggest that this
is the case only for light targets, as C, Al or Fe, while an excess
of yield is obtained for Au~\cite{danielaPavia}.
To study the origin of the rescattered strength,
the calculations were repeated by neglecting the mean field
part $S^h_{MF}$ in
Eq.~(\ref{eq:Shtotal}).
 In the case of ${}^{12}{\rm C}$ no appreciable difference was
found for missing momenta above 400~MeV, even in perpendicular
kinematics. This shows that the rescattered strength seen
experimentally at these kinematics originates only from the
correlated region itself.
The situation is different in the case of Au since for heavy nuclei the 
mean field spectral function extends to very high missing energies (up
to $\sim$100~MeV) and can therefore contribute to the experimental yield.
Figure~\ref{fig:res5} compares the theoretical reduced spectral strength
of gold with the analogous result obtained
when the effects of $S^h_{MF}$ are neglected. As one can see,
relevant mean field
contributions appear already in parallel kinematics for momenta
below 500~MeV/c, while the rescattering
at high missing momenta is still dominated by the correlated part $S^h_{corr}$.
One should note that since little reliable experimental information for
$S^h_{corr}$ is available to date, the correlated strength can be extracted
from the experimental data only in a self-consistent fashion.
This of course requires a proper treatment of the FSI.
For heavy nuclei a proper description of the mean field strength is also
relevant and could explain the increase of rescattering with respect
to $A$ seen experimentally.


\section{Conclusions}

 This contribution reports about an ongoing work aimed to study the
effects of final state interactions in $(e,e'p)$ reactions, as
generated by rescattering effects.
 The two-step rescattering processes that involve the intermediate
propagation of a nucleon have been approached by using a semiclassical
model and
preliminary calculations have been reported for ${}^{12}{\rm C}(e,e'p)$
and ${}^{197}{\rm Au}(e,e'p)$.
It is seen that the contributions from final state interactions
increases with the mass number $A$ of the target. Moreover, for a given
nucleus they are found to be much smaller in parallel kinematics
than in perpendicular ones.
In the latter case a large
amount of strength is shifted from regions where the spectral function
is big to regions where it is smaller, thus overwhelming the experimental
yield from the direct process.
This confirms the studies of Ref.~\cite{E97proposal}.

At $E_m >250$MeV the present experimental results in parallel
kinematics exceed the calculated direct 
plus rescattering contributions by about an order of magnitude.
 This discrepancy is visible even at lower missing energies when
perpendicular kinematics are considered.
 This is 
presumably due to the excitation of $\Delta$ resonances which requires 
higher energies, about $\sim$~300~MeV more. The subsequent decay of
the $\Delta$ can shift a sizable amount of strength toward lower missing
energies, eventually affecting the measurements in the correlated region.
 The inclusion of these effects in the present model 
will be the topic of future work.

\section*{acknowledgments}

The authors would like to thank D. Rohe for several useful discussions
and for providing the data of the E97-006 experiment.
This work is supported by the Natural
Sciences and Engineering Research Council of Canada (NSERC) and
by the ``Stichting voor Fundamenteel Onderzoek der Materie (FOM)'',
which is financially supported by the ``Nederlandse Organisatie voor
Wetenschappelijk Onderzoek (NWO)''.

\end{document}